\documentstyle[aps,amssymb,preprint,prd]{revtex}
\newcommand{\be}{\begin{equation}}
\newcommand{\ee}{\end{equation}}
\newcommand{\bea}{\begin{eqnarray}}
\newcommand{\nn}{\nonumber}
\newcommand{\eea}{\end{eqnarray}}

\begin{document}

\draft
\title{\textbf Casimir Effect in 2D Stringy Black Hole Backgrounds}

\author{T. Christodoulakis\footnote{Electronic address: tchris@cc.uoa.gr}, 
G.A. Diamandis\footnote{Electronic address: gdiamant@cc.uoa.gr}, 
B.C. Georgalas\footnote{Electronic address: vgeorgal@cc.uoa.gr} 
and E.C. Vagenas\footnote{Electronic address: hvagenas@cc.uoa.gr}}
\address{Department of Physics, University of Athens, Panepistimioupolis, Ilisia 157 71 Athens, Greece}
\date{\today}
\maketitle

\begin{abstract}
 We consider the two-dimensional ``Schwarzschild" and
``Reissner-Nordstr\"{o}m'' stringy black holes as systems of Casimir
type. We explicitly calculate the energy-momentum tensor of a
massless scalar field satisfying Dirichlet boundary conditions on
two one-dimensional ``walls''. These results are obtained using
the Wald's axioms. Thermodynamical quantities such as pressure,
specific heat, isothermal compressibility and entropy of the
two-dimensional stringy black holes are calculated. A comparison
is made between the obtained results and the laws of
thermodynamics. The results obtained for the extremal (Q=M)
stringy two-dimensional charged black hole are identical in all three different
 vacua used; a fact that indicates its quantum stability.
\end{abstract}

\section{Introduction}
 A significant tool for investigating the thermodymamical
 properties of black holes is the Casimir effect. As it is well
 known in 1948 H.B.G. Casimir \cite{casimir} evaluated the
 electromagnetic energy
 localized between two conducting plates. The disturbance to
 the electromagnetic vacuum induced by the two parallel plates
 is actually observable.
 The so called Casimir effect \cite{mostepanenko}  is viewed as a tractable model
 of field theoretical effects associated with the geometry of space
\cite{emilio1,emilio2,emilio3}.
 \par\noindent
 In order to examine the analogous effects for non-trivial
 gravitational backgrounds we need the vacuum expectation values
 of the energy-momentum tensor. There are
many procedures  \cite{birrell,fulling,bryce,vassilevich}
 for calculating the vacuum expectation value of
the energy-momentum tensor such as the dimensional regularization
\cite{capper1,capper2,deser}, Green's function method
\cite{plunien,candelas}, heat kernel method \cite{greiner,gilkey},
zeta function regularization \cite{hawking1},
point-splitting method \cite{davies1,davies2,christ},
Pauli-Vilars regularization \cite{vilenkin}.
\par
\noindent
We restrict the form of the
renormalized energy-momentum tensor of a massless scalar field \cite{christensen}
(without employing the full theory of regularization) by using the
trace of $T_{\mu\nu}$ and enforcing Wald's axioms \cite{wald1,wald11}
which are :
\begin{enumerate}
\item The expectation values of the energy-momentum tensor are covariantly conserved.
\item Causality holds.
\item In Minkowski spacetime, standard results should be obtained.
\item Standard results for the off-diagonal elements should also be obtained.
\item The energy-momentum tensor is a local functional of the metric, i.e. it depends only on
the metric and its derivatives which appear through the Riemann curvature tensor and
 the metric's covariant derivatives up to second order.
\end{enumerate}
\noindent
In working this procedure a detailed expression for the renormalized
energy-momentum tensor is obtained once the stringy
two-dimensional ``Schwarzschild" (massive)  \cite{witten,mandal}
and ``Reissner-Nordstr\"{o}m'' (charged)  \cite{lee,nappi} black hole backgrounds
 are treated as systems of Casimir type \cite
{birrell,setare1,setare2}.
\par
\noindent The outline of this paper is as follows. In Section II
and III the vacuum expectation value of the energy-momentum tensor
is explicitly evaluated for the above mentioned stringy black hole
backgrounds, respectively, in the Boulware vacuum (labeled by
$\eta$) \cite{boulware}, the Hartle-Hawking vacuum (labeled by
$\upsilon$) \cite{hartle,gibbons,israel}
 and the Unruh vacuum (labeled by $\xi$)
\cite{unruh}. The energy density, pressure, energy and the corresponding
force between the two ``Dirichlet walls'' are calculated asymptotically. The
thermodynamical quantities specific heat, thermal compressibility
and entropy exhibit a fictitious violation of the second
thermodynamical law. In Section IV the results are discussed and
conclusions are given.

\section{``Schwarzschild" Black Hole}
The line element of the stringy two-dimensional ``Schwarzschild'' black
hole \cite{cghs} which is a low-energy solution of an effective string action
\cite{witten,mandal} is given as :
\be
ds^2=-g(r)dt^{2} +\frac{dr^{2}}{g(r)}
\label{element1}
\ee
where the metric function is :
\be
g(r)=1-\frac{M}{\lambda}e^{-2\lambda r} \label{metric1} \ee the
radial coordinate take values $r_{H}<r<+\infty$ and the event
horizon $\mathcal{H}$ is placed at the point :
\be
r_{H}=\frac{1}{2\lambda}ln\left(\frac{M}{\lambda}\right)
\label{root1}.
\ee
The line element (\ref{element1}) is written in ``Schwarzschild''
gauge.
\newline
In the conformal gauge which we are going to use in our
calculations the line element is given by :
\be
ds^2=\Omega(x)\left(-dt^2 + dx^2\right)
\ee
the conformal factor is :
\be
\Omega(x)=\frac{1}{1+e^{-2\lambda x}}
\label{conformal1}
\ee
where the conformal variable :
\be
x=\frac{1}{2\lambda}ln\left[e^{2\lambda(r-r_H)}-1\right]
\ee
takes values $-\infty<x<+\infty$
and the corresponding conformal factor takes values :
\be
0<\Omega(x)<1 .
\ee
The non-zero Christoffel symbols are :
\be
\Gamma^{t}_{xt}=\Gamma^{x}_{xx}=\Gamma^{x}_{tt}=
\frac{1}{2\Omega(x)}\frac{d\Omega(x)}{dx}=
\lambda \left[\frac{e^{-2\lambda x}}
{1+e^{-2\lambda x}}\right] \label{gamma1}.
 \ee
 The Ricci scalar is given as :
\be
R(x)=4\lambda^{2}
\left[\frac{e^{-2\lambda x}}{1+e^{-2\lambda x}}\right]. \ee It is
well known that the trace $T^{\alpha}_{\alpha}(x)$ of the
energy-momentum tensor vanishes classically for a conformally
invariant theory. However in the semiclassical approximation which
is the case to be discussed here the trace is nonzero in the
regularization process and specifically in two dimensions is give
by \cite{capper1,capper2,christensen,capri} :
\be
T^{\alpha}_{\alpha}(x)=\frac{R(x)}{24\pi} .
 \ee
 Thus for the
two-dimensional ``Schwarzschild'' black hole background
(\ref{element1})-(\ref{root1}) the trace of the energy-momentum
tensor is :
\be
T^{\alpha}_{\alpha}(x)=
\frac{\lambda^{2}}{6\pi}
\left[\frac{e^{-2\lambda x}}{1+e^{-2\lambda x}}\right] .
\label{trace1} \ee Applying  Wald's first axiom, the conservation
equation must be fulfilled by the regularized expectation value of
the energy-momentum tensor $<T^{\mu}_{\quad \nu}>_{reg} \equiv
T^{\mu}_{\quad \nu}$ :
\be
T^{\mu} _{\quad \nu;\mu}=0 \label{conservation} \ee
which ``splits'' in two equations : \bea
\frac{dT^{x}_{t}}{dx}+\Gamma^{t}_{tx}T^{x}_{t}
-\Gamma^{x}_{tt}T^{t}_{x}=0\\
\frac{dT^{x}_{x}}{dx}+\Gamma^{t}_{tx}T^{x}_{x}
-\Gamma^{t}_{tx}T^{t}_{t}=0 \eea and since $T^{t}_{x}=-T^{x}_{t}$
and $T^{t}_{t}=T^{\alpha}_{\alpha}-T^{x}_{x}$, we get : \bea
\frac{dT^{x}_{t}}{dx}+2\Gamma^{t}_{tx}T^{x}_{t}=0 \label{em1}\\
\frac{dT^{x}_{x}}{dx}+2\Gamma^{t}_{tx}T^{x}_{x}=
\Gamma^{t}_{tx}T^{\alpha}_{\alpha} . \label{em2} \eea Substituting
the Christoffel symbols (\ref{gamma1}) into (\ref{em1}), (\ref{em2})
and solving them, we get respectively :
\be
T^{x}_{t}(x)=\alpha\Omega^{-1}(x)=\alpha\left(1+e^{-2\lambda
x}\right) \ee
\be
T^{x}_{x}(x)=\Omega^{-1}(x)\left[H_{2}(x)+\beta\right]
\ee
where
\be
H_{2}(x)=\frac{1}{2}\int\limits^{x}_{x_{H}}
\frac{d\Omega(x')}{dx'}T^{\alpha}_{\alpha}(x')dx'
\label{rate}
\ee and the parameters $\alpha$, $\beta$ are constants of integration while
the point $x_{H}$ is where the event horizon $\mathcal{H}$ is
placed.
It can be shown that $H_{2}(x)$ for the stringy two-dimensional black
hole background (\ref{element1}-\ref{conformal1}) becomes :
\be
H_{2}(x)=\frac{\lambda^{2}}{24\pi}-\frac{\lambda^{2}}{24\pi}
\left(1-\Omega(x)\right)^2 \label{H21}. \ee Now the following
limiting values of $H_{2}(x)$ from (\ref{H21}) are obtained :
 \begin{eqnarray*}
\begin{array}{ll}
 if\hspace{0.5cm}
x \rightarrow -\infty \hspace{0.2cm}(r=r_{H})& then\hspace{0.5cm}
H_{2}(x)= 0 \hspace{0.2cm}\left(\Omega(x)=0\right) \nn
\\ if\hspace{0.5cm} x \rightarrow +\infty \hspace{0.2cm} (r\rightarrow +\infty) & then\hspace{0.5cm}
H_{2}(x)=\frac{\lambda^{2}}{24\pi}\hspace{0.2cm}\left(\Omega(x)=1\right)\nn.
\end{array}
\end{eqnarray*}
\noindent Keeping in mind that for any two-dimensional
background the most general expression of the regularized
energy-momentum tensor is :
\be
T^{\mu}_{\nu}=\left[ \begin{array}{cc}
 T^{\alpha}_{\alpha}(x)- \Omega^{-1}(x)H_{2}(x) & 0 \\
 0 &  \Omega^{-1}(x)H_{2}(x)
\end{array}
\right]+\Omega^{-1}(x)
\left[ \begin{array}{cc}
-\beta & -\alpha\\
\alpha & \beta
\end{array}
\right]
\label{generalT} \ee we obtain : \bea T^{\mu}_{\nu}= & \left[
\begin{array}{cc}
 \frac{\lambda^{2}}{6\pi}
\left[1-\Omega(x)\right] - \Omega^{-1}(x)
\left(\frac{\lambda^{2}}{24\pi}-\frac{\lambda^{2}}{24\pi}
\left[1-\Omega(x)\right]^2\right)
& 0 \\
 0 &  \Omega^{-1}(x)\left(
\frac{\lambda^{2}}{24\pi}-\frac{\lambda^{2}}{24\pi}
\left[1-\Omega(x)\right]^2\right)
\end{array} \right] & \nn\\
 & +\hspace{0.5cm}\Omega^{-1}(x)
\left[ \begin{array}{cc}
-\beta & -\alpha\\
\alpha & \beta
\end{array}
\right]
 \label{em3} \eea where the stringy background
(\ref{element1})-(\ref{root1}) and relations (\ref{trace1}),
(\ref{H21}) have been used. In this expression the only unknowns
are the parameters $\alpha$ and $\beta$; we hope to determine them
imposing the third Wald's axiom treating the two-dimensional
``Schwarzschild'' black hole as a Casimir system \cite{birrell}.
Two one-dimensional ``walls" at a proper distance $L$ (between them)
are placed at arbitrary points $x_1$ and $x_2$. The massless scalar
field whose energy-momentum tensor we try to evaluate satisfies
the Dirichlet boundary conditions on the ``walls'', i.e.
$\phi(x_{1})=\phi(x_{2})=0$.
\par
\noindent
The standard Casimir energy-momentum tensor in the Minkowski
spacetime is already known \cite{birrell,fulling} :
\be
T^{\mu}_{\nu}=\frac{\pi}{24L^{2}}
\left[\begin{array}{cc}
-1 & 0 \\
0 & 1
\end{array}
\right]. \label{emb} \ee The two-dimensional ``Schwarzschild''
black hole is asymptotically flat, i.e. at infinity (towards
$\mathcal{J}^+$) is Minkowski spacetime, so the constants of
integration $\alpha$ and $\beta$ are evaluated demanding the
regularized energy-momentum tensor given in (\ref{em3}) to
coincide with the standard Casimir energy-momentum tensor
(\ref{emb}) at infinity, i.e. $x\rightarrow +\infty$, or
equivalently setting $\Omega(x)=1$.
\par
\noindent
Therefore we get :
\be
\beta=\frac{\pi}{24L^{2}}-\frac{\lambda^{2}}{24\pi}\\
\hspace{1cm}\alpha =0 \ee and the regularized energy-momentum
tensor has been explicitly calculated :
\bea T^{(\eta)\mu}_{\nu}=
& \left[
\begin{array}{cc}
 \frac{\lambda^{2}}{6\pi}
\left[1-\Omega(x)\right] - \Omega^{-1}(x)
\left(\frac{\lambda^{2}}{24\pi}-\frac{\lambda^{2}}{24\pi}
\left[1-\Omega(x)\right]^2\right)
& 0 \\
 0 &  \Omega^{-1}(x)\left(
\frac{\lambda^{2}}{24\pi}-\frac{\lambda^{2}}{24\pi}
\left[1-\Omega(x)\right]^2\right)
\end{array} \right] & \nn\\
 & +\hspace{0.5cm}\Omega^{-1}(x)\left(\frac{\pi}
{24L^{2}}-\frac{\lambda^{2}}{24\pi} \right)
\left[ \begin{array}{cc}
-1 & 0\\
0 & 1
\end{array}
\right] \label{em4} \eea where $\eta$ denotes that the regularized
energy-momentum tensor has been calculated under the assumption
that there are no particles (vacuum state) at infinity (Boulware
vacuum). Thus we have obtained the regularized energy-momentum
tensor $T^{(\eta)\mu}_{\nu}$ as a direct sum :
\be
T^{(\eta)\mu}_{\nu}=T^{\mu}_{\nu(gravitational)}+T^{\mu}_{\nu(boundary)}
\ee where the first term denotes the contribution to the vacuum
polarization due to the non-trivial topology in which the
contribution of the trace anomaly is included and the second term
denotes the contribution due to the presence of the two ``Dirichlet walls''.
\par
\noindent
In the Boulware vacuum the detected negative
energy density ${\rho}$ will asymptotically ($x\rightarrow +\infty$, or
equivalently setting $\Omega(x)=1$) be :
\be
\rho=T^{(\eta)t}_{t}=-\frac{\pi}{24L^{2}}
\label{density1}
\ee the detected pressure $p$ is asymptotically :
\be
p=-T^{(\eta)x}_{x}=-\frac{\pi}{24L^{2}}
\ee
the detected negative energy $E$ will asymptotically be :
\be
E(L)=\int^{L}_{0}\rho dx =-\frac{\pi}{24L}\label{energy1} .\ee
 The corresponding
force $F$ (i.e. the force due to the localized energy between
the``walls") will be attractive as expected:
\be
F(L)=-\frac{\partial E(L)}{\partial L}=-\frac{\pi}{24L^{2}}< 0 .
\ee \par \noindent Let us stress that the above formula (\ref{energy1}), and the
ones that follow regarding the different vacua, consider both the
``walls" at the asymptotic region. If one wishes to include more 
specific effects of the background one has to place them  
at arbitrary points $x_1$ and $x_1 +L$. Then the corresponding energy will be :
\bea
 E(x_1,L)&=&\int^{x_1 +
L}_{x_1} T^{t}_{t}(x)dx\nn\\
 &=& -\frac{1}{8\lambda}
\left(\frac{\pi}{6L^2}-\frac{\lambda^2}{6\pi}\right)e^{-2 \lambda
x_1}-\frac{1}{8\lambda}\left(\frac{\lambda^2}{6\pi}-\frac{\pi}{6L^2}\right)e^{-2\lambda
L }e^{-2 \lambda x_1}\nn\\ &&-\frac{\pi}{24
L}+\frac{\lambda}{16\pi}\log[1+e^{-2 \lambda
x_1}]-\frac{\lambda}{16\pi}\log[1+e^{-2\lambda L }e^{-2 \lambda
x_1}]
 \label{energy11}
\eea
 yielding (\ref{energy1}) as $ x_1 \rightarrow +\infty $.
\par
\noindent In the Hartle-Hawking vacuum the black hole is in
thermal equilibrium with an infinite reservoir of black body
radiation at temperature $T$ and the standard Casimir
energy-momentum tensor (\ref{emb}) is modified by an additional
term \cite{christensen} :
\be
T^{\mu}_{\nu}=\frac{\pi T^{2}}{12}
\left[
\begin{array}{cc}
-2 & 0\\
0 & 2
\end{array}
\right]=\frac{\pi T^{2}}{6}
\left[
\begin{array}{cc}
-1 & 0\\
0 & 1
\end{array}
\right ] \label{embath}. \ee Setting $T$ equal to the Hawking
temperature of the two-dimensional ``Schwarzschild'' black hole,
i.e. $T_{H}=\frac{\lambda}{2\pi}$, the Casimir energy-momentum
tensor becomes :
\be
T^{\mu}_{\nu}=\frac{\pi}{24L^{2}}
\left[\begin{array}{cc}
-1 & 0 \\
0 & 1
\end{array}
\right]+
\frac{\lambda^{2}}{24\pi}
\left[
\begin{array}{cc}
-1 & 0\\
0 & 1
\end{array}
\right] \label{emb+bath}. \ee Therefore we now obtain :
\be
\beta=\frac{\pi}{24L^{2}}\\ \hspace{1cm}\alpha =0 \ee and the
regularized energy-momentum tensor becomes : \bea
T^{(\upsilon)\mu}_{\nu}= & \left[ \begin{array}{cc}
 \frac{\lambda^{2}}{6\pi}
\left[1-\Omega(x)\right] - \Omega^{-1}(x)
\left(\frac{\lambda^{2}}{24\pi}-\frac{\lambda^{2}}{24\pi}
\left[1-\Omega(x)\right]^2\right)
& 0 \\
 0 &  \Omega^{-1}(x)\left(
\frac{\lambda^{2}}{24\pi}-\frac{\lambda^{2}}{24\pi}
\left[1-\Omega(x)\right]^2\right)
\end{array} \right] & \nn\\
 & +\hspace{0.5cm}\Omega^{-1}(x)\left(\frac{\pi}
{24L^{2}}\right)
\left[ \begin{array}{cc}
-1 & 0\\
0 & 1
\end{array}
\right]
\label{em5}
\eea
where $\upsilon$ denotes that the regularized energy-momentum
tensor has been been calculated under the assumption
that massless particles (black body radiation) are detected at
infinity (towards $\mathcal{J}^+$) (Hartle-Hawking vacuum).
Thus the regularized energy-momentum
tensor $T^{(\upsilon)\mu}_{\nu}$ is :
\be
T^{(\upsilon)\mu}_{\nu}=T^{\mu}_{\nu(gravitational)}+T^{\mu}_{\nu(boundary)}
+T^{\mu}_{\nu(bath)} \ee where the last term denotes the
contribution to the vacuum polarization due to thermal bath at
temperature $T_{H}$.
\par
\noindent In this vacuum the asymptotically 
($x\rightarrow +\infty$, or
equivalently setting $\Omega(x)=1$) detected energy density and pressure
are : \bea \rho=T^{(\upsilon)t}_{t}=-\left(\frac{\pi}{24L^{2}}
+\frac{\lambda^{2}}{24\pi}\right)
\\
p=-T^{(\upsilon)x}_{x}=-\left(\frac{\pi}{24L^{2}}
+\frac{\lambda^{2}}{24\pi}\right) \eea likewise the detected negative energy is:
\be
E(L, T_{H})=\int^{L}_{0}\rho dx = -\left( \frac{\pi}{24L}+
\frac{\lambda^{2}}{24\pi}L \right)=-\left( \frac{\pi}{24L}+
\frac{\pi L}{6}T^{2}_{H} \right)
 \label{energyHH1} \ee and the corresponding
force $F$ between the ``walls'' is not always attractive :
\be
F(L, T_{H})=-\left(\frac{\partial E(L, T_{H})}{\partial
L}\right)_{T_{H}}=-\frac{\pi}{24L^{2}}+
\frac{\lambda^{2}}{24\pi}=-\frac{\pi}{24L^{2}}+
\frac{\pi}{6}T^{2}_{H} .
 \ee
 It is clear that the corresponding force is :
\begin{description}
\item \hspace{2cm}{\bf (a) attractive}
\be L < \frac{1}{2\hspace{0.1cm} T_{H}}=\frac{\pi}{\lambda} \ee
\item \hspace{2cm}\bf (b) zero \be L = \frac{1}{2\hspace{0.1cm}
 T_{H}} \ee
\item \hspace{2cm}\bf (c) repulsive
\be L > \frac{1}{2\hspace{0.1cm} T_{H}} \ee
\end{description}
Thus if the last condition is satisfied the outer ``wall'' moves
towards infinity. It can be studied as a ``moving mirror''
creating particles. The energy rate detected at infinity can be given
by the second term in equation (\ref{energyHH1}) :
\be
\frac{dE}{dt}=\frac{\lambda^{2}}{24\pi}L =\frac{\pi L}{6}T^{2}_{H}
\ee and this is (for the massless two-dimensional field) the
rate at which the energy is radiated \cite{cghs,emparan1,emparan2}.\par
\noindent
The appearance of the repulsive nature of the corresponding force will be discussed
in Section IV.
\par
\noindent In the Unruh vacuum an outward flux of radiation is
detected at infinity (towards $\mathcal{J}^+$). Since the
two-dimensional ``Schwarzschild'' black hole has been proven to
radiate and its spectrum distribution is purely thermal at the
Hawking temperature $T_{H}$ \cite{hawking2,giddings}, the Unruh
vacuum state can be identified with the vacuum obtained after the
two-dimensional ``Schwarzschild'' black hole has settled down to
an ``equilibrium'' of temperature $T_{H}$. The standard Casimir
energy-momentum tensor (\ref{emb}) will be modified by an
additional term \cite{christensen} :
\be
T^{\mu}_{\nu}=\frac{\pi T_{H}^{2}}{12}
\left[
\begin{array}{cc}
-1 & 1\\
1 & 1
\end{array}
\right]= \frac{\lambda^{2}}{48\pi}\left[
\begin{array}{cc}
-1 & 1\\ 1 & 1
\end{array}
\right] \label{emrad}. \ee The Casimir energy-momentum tensor is now given by :
\be
T^{\mu}_{\nu}=\frac{\pi}{24L^{2}}
\left[\begin{array}{cc}
-1 & 0 \\
0 & 1
\end{array}
\right]+
\frac{\lambda^{2}}{48\pi}
\left[
\begin{array}{cc}
-1 & -1\\
1 & 1
\end{array}
\right]
\label{emb+rad}.
\ee
Therefore we get :
\be
\beta=\frac{\pi}{24L^{2}}-\frac{\lambda^{2}}{48\pi}\\
\hspace{1cm}\alpha =\frac{\lambda^{2}}{48\pi} \ee and the
regularized energy-momentum tensor becomes : \bea
T^{(\xi)\mu}_{\nu}= & \left[ \begin{array}{cc}
 \frac{\lambda^{2}}{6\pi}
\left[1-\Omega(x)\right] - \Omega^{-1}(x)
\left(\frac{\lambda^{2}}{24\pi}-\frac{\lambda^{2}}{24\pi}
\left[1-\Omega(x)\right]^2\right)
& 0 \\
 0 &  \Omega^{-1}(x)\left(
\frac{\lambda^{2}}{24\pi}-\frac{\lambda^{2}}{24\pi}
\left[1-\Omega(x)\right]^2\right)
\end{array} \right] & \nn\\
 & +\hspace{0.5cm}\Omega^{-1}(x)
\left[ \begin{array}{cc}
-\frac{\pi}{24L^{2}}+\frac{\lambda^{2}}{48\pi} &
-\frac{\lambda^{2}}{48\pi}  \\
\frac{\lambda^{2}}{48\pi}  &
\frac{\pi}{24L^{2}}-\frac{\lambda^{2}}{48\pi}
\end{array}
\right]
\label{em6}
\eea
where $\xi$ denotes that the regularized energy-momentum
tensor has been been calculated under the assumption
that massless particles are detected at infinity due to
the Hawking radiation of the two-dimensional ``Schwarzschild''
black hole. Thus the regularized energy-momentum
tensor $T^{(\xi)\mu}_{\nu}$ becomes :
\be
T^{(\xi)\mu}_{\nu}=T^{\mu}_{\nu(gravitational)}+T^{\mu}_{\nu(boundary)}
+T^{\mu}_{\nu(radiation)}
\ee
where the last term denotes the contribution
to the vacuum polarization due to Hawking radiation at temperature
$T_{H}$.
\par
\noindent In this vacuum the detected energy density and pressure
are asymptotically ($x\rightarrow +\infty$, or
equivalently setting $\Omega(x)=1$) :
 \bea \rho=T^{(\xi)t}_{t}=-\left(\frac{\pi}{24L^{2}}
+\frac{\lambda^{2}}{48\pi}\right)
\\
p=-T^{(\xi)x}_{x}=-\left(\frac{\pi}{24L^{2}}
+\frac{\lambda^{2}}{48\pi}\right) \eea 
likewise for the detected
negative energy :
\be
E(L, T_{H})=\int^{L}_{0}\rho dx = -\left( \frac{\pi}{24L}+
\frac{\lambda^{2}}{48\pi}L \right) =
 -\left( \frac{\pi}{24L}+
\frac{\pi L}{12}T^{2}_{H}\right)
 \label{energyU1} \ee and the corresponding force $F$ between the
``walls'' is not always attractive :
\be
F(L, T_{H})=-\left(\frac{\partial E(L, T{H})}{\partial
L}\right)_{T_{H}}=-\frac{\pi}{24L^{2}}+
\frac{\lambda^{2}}{48\pi}=-\frac{\pi}{24L^{2}}+
\frac{\pi}{12}T^{2}_{H} .
 \ee The corresponding force will now be :
\begin{description}
\item \hspace{2cm}{\bf (a) attractive}
\be L < \frac{1}{\sqrt{2} \hspace{0.1cm}
T_{H}}=\sqrt{2}\frac{\pi}{\lambda} \ee
\item \hspace{2cm}\bf (b) zero \be L = \frac{1}{\sqrt{2}\hspace{0.1cm}
 T_{H}} \ee
\item \hspace{2cm}\bf (c) repulsive
\be L > \frac{1}{\sqrt{2}\hspace{0.1cm} T_{H}} . \ee
\end{description}
Thus if the outer ``wall'' is placed at a distance $L$ such that
the last condition is satisfied then it will move towards
infinity. It can be studied as a ``moving mirror'' creating
particles whose energy rate detected at infinity is given by the
second term in equation (\ref{energyU1}):
\be
\frac{dE}{dt}=\frac{\lambda^{2}}{48\pi}L=\frac{\pi
L}{12}T^{2}_{H}. \ee Thus as before this is for the massless
two-dimensional field the rate at which
 the energy is radiated \cite{cghs,emparan1,emparan2}.
\par\noindent
In this vacuum it will be interesting to calculate some
thermodynamical quantities and to consider these results
with respect to the laws of thermodynamics.
\par \noindent
The specific heat is given as :
\be
C_{V}=-\frac{\partial E}{\partial T} =-\left(\frac{\pi
L}{6}\right) T_{H}= -\left(\frac{L}{12}\right)\lambda \ee with $V$
the ``volume'' between the two ``walls'', i.e. the distance $L$
in this case, and the isothermal compressibility is :
\be
\kappa_{T}=-\frac{1}{L}\left( \frac{\partial L}{\partial
p}\right)_{T_{H}} =-\left(\frac{12}{\pi}\right)L^{2} . \ee
 The negative values of the specific heat and the isothermal
compressibility are a violation of the second law of
thermodynamics which requires $C_{V}\geq 0$ and $\kappa_{T} \geq
0$ \cite{sassaroli}.
 This thermodynamical instability at least for $C_{V}$ is a common
feature in black hole physics, using the semiclassical approximation.
This may be resolved by a more complete quantum treatment
and the inclusion of back reaction effects \cite{georgalas}.
\par
\noindent The entropy of the stringy two-dimensional
``Schwarzschild'' black hole seen as a Casimir system is given
(applying the first thermodynamical law) by :
\be
S_{Casimir}=S_{\left(T=0 \right)}-\left(\frac{\pi L}{6}\right)
 T_{H}=
S_{\left(T=0 \right)}-\left(\frac{L}{12}\right)\lambda \ee and
according to the third law of thermodynamics : \be S \rightarrow
0\hspace{0.5cm}as\hspace{0.5cm}T \rightarrow 0 \ee the entropy is
:
\be
S_{Casimir}=-\left(\frac{\pi L}{6}\right) T_{H}
=-\left(\frac{L}{12}\right)\lambda .
\label{entropy1}
\ee
The entropy calculated here seems to violate of the second
thermodynamical law. This is not true since the entropy calculated
here has not been obtained through a statistical counting of
microstates. The expression (\ref{entropy1}) is the part of the
thermodynamical entropy $S^{TM}$ due to the vacuum polarization
(virtual particles)-it doesn't need to have a statistical interpretation-
and so it is not forbidden to be negative \cite{zaslavskii}.
\par \noindent
Being more precise the
thermodynamical entropy which gives the contribution of
quantum fields (radiation and massive fields) is given as :
\be
S^{TM}=S^{SM}+S_{0} \ee where $S^{SM}$ is the
statistical-mechanical part of the entropy (statistical counting
of microstates) which is absent here and in our case
$S_{0}=S_{Casimir}$ is a quantity which gives the contribution of
the vacuum polarization.
\par
\noindent Finally the entropy of the two-dimensional
``Schwarzschild black hole is :
\be
S_{bh}=S_{classical}+S^{TM}
\ee
where $S_{classical}$ is the entropy from the classical
gravitational action and the thermodynamical entropy
$S^{TM}$ is the one-loop quantum correction \cite{solodukhin}.
\section{``Reissner-Nordstr\"{o}m'' Black Hole}
The line element of the stringy two-dimensional
``Reissner-Nordstr\"{o}m'' (charged)
 black hole \cite{lee,nappi} in the Schwarzschild gauge is given by :
\be
ds^2 = -g(r)dt^2 + g^{-1}(r)dr^2
\label{element2}
\ee
where
\be
g(r) = 1 - \frac{M}{\lambda} e^{-2\lambda r} +
\frac{Q^2}{4 \lambda ^2} e^{-4\lambda r}
\label{metric2}
\ee
with $0<t<+\infty$, $r_+<r<+\infty$, $r_+$ being the
future event horizon of the black hole.
\newline
Following a parametrization analogous
to the four-dimensional case the metric function factorizes as :
\be
g(r)=(1-\rho_- e^{-2\lambda r})(1-\rho_+e^{-2\lambda r})
\ee
where
\be
\rho_\pm=\frac{M}{2 \lambda}\pm\frac{1}{2\lambda} \sqrt{M^2-Q^2}
\label{roots2} \ee we can recognize immediately the ``outer'' event
horizon $\mathcal{H}^+$ placed at the point
$r_+=\frac{1}{2\lambda}ln\rho_+$, while the ``inner" horizon
$\mathcal{H}^-$ is at the point $r_-=\frac{1}{2\lambda}ln\rho_-$.
\newline
In the extremal case ($Q=M$)
the two surfaces coincide in a single event horizon
at the point :
\be
r_H=\frac{1}{2 \lambda}ln \left( \frac{M}{2 \lambda} \right) .
\ee
The line element (\ref{element2})-(\ref{metric2})
in conformal gauge is written :
\be
ds^2=\Omega(x)\left(-dt^2 + dx^2\right)
\ee
with the conformal factor :
\be
\Omega(x)=\frac{X(x)\left(X(x)+\mu\right)}{\left(X(x)+1\right)^2}
\label{conformal2} \ee
\be
0<\Omega(x)<1
\ee
where the conformal variable :
\be
x=\frac{1}{2\lambda \mu}ln\left[X \left(X+\mu
\right)^{\mu-1}\right] \ee with $-\infty<x<+\infty$ and the
asymmetric variable $X$ is :
\be
X=e^{2\lambda(r-r_{+})}-1=\frac{e^{2\lambda r}}{\rho_{+}}-1 \ee
\be
0<X<+\infty .
 \ee The new parameter $\mu$ is given by :
\be \mu =1- \frac{\rho_{-}}{\rho_{+}} . \ee The non-zero
Christoffel symbols are : \be
\Gamma^{t}_{xt}=\Gamma^{x}_{xx}=\Gamma^{x}_{tt}=
\frac{1}{2\Omega(x)}\frac{d\Omega(x)}{dx}=\lambda
\frac{\left[\left(X+\mu \right)+X \left(1-\mu \right)\right]}
{\left(1+X\right)^{2}} . \ee The Ricci scalar is given as : \be
R(x)=4\lambda^{2}\left[\frac{2\left(X(x)+\mu
-1\right)+\mu(1-X(x))}{\left(X(x)+1 \right)^{2}}\right] \ee and in
the semiclassical approximation the trace (or conformal) anomaly
for the stringy two-dimensional background
(\ref{element2})-(\ref{roots2}) is : \be
T^{\alpha}_{\alpha}(x)=\frac{\lambda^{2}}{6\pi}
\left[\frac{2\left(X(x)+\mu -1\right)+\mu(1-X(x))}{\left(X(x)+1
\right)^{2}}\right]. \label{trace2} \ee The conservation equation
(\ref{conservation}) again ``splits'' in two equations : \be
\frac{d}{dx}\left[\Omega(x)T^{x}_{t}\right]=0 \ee \be
\frac{d}{dx}\left[\Omega(x)T^{x}_{x}\right]=\frac{1}{2}
\left(\frac{d\Omega(x)}{dx}\right)T^{\alpha}_{\alpha}(x)
 \ee
and by integration we obtain :
\be
T^{x}_{t}(x)=\alpha\Omega^{-1}(x)= \alpha
\frac{\left(X(x)+1\right)^2} {X(x)\left(X(x)+\mu\right)} \ee
\be
T^{x}_{x}(x)=\Omega^{-1}(x)\left[H_{2}(x)+\beta\right]=
\frac{\left(X(x)+1\right)^2}
{X(x)\left(X(x)+\mu\right)}\left[H_{2}(x)+\beta\right]
\ee
using (\ref{rate}) $H_{2}(x)$ is now :
\be
H_{2}(x)=\frac{\lambda^{2}}{12\pi}\int^{X(x)}_{X(x_{+})}
\frac{\left[\left(X+\mu\right)+X(1-\mu)\right] \left[2\left(X+\mu
-1\right)+\mu(1-X)\right]} {\left(1+X \right)^{5}}dX \ee and the
parameters $\alpha$, $\beta$ are constants of integration while the
point $x_{+}$ is where the ``outer'' event horizon $\mathcal{H}^{+}$
is placed.
\newline
Thus the quantity $H_{2}(x)$ becomes :
\be
H_{2}(x)=\frac{\lambda^{2}}{24\pi} \left[\mu ^{2}+H_{1}(x)\right]
\label{H22} \ee with
\be
H_{1}(x)=-\frac{4(1-2\mu+\mu^{2})}{\left(1+X\right)^{4}}
+\frac{4(2-3\mu+\mu^{2})}{\left(1+X\right)^{3}}
-\frac{(\mu^{2}-4\mu+4)}{\left(1+X\right)^{2}}
\label{H1}
\ee
with the following limiting values :
\begin{eqnarray*}
\begin{array}{ll}
if\hspace{0.5cm} x \rightarrow -\infty
\hspace{0.2cm} \left(X \rightarrow 0\right)
&then\hspace{0.5cm} H_{2}(x)= 0
\hspace{0.2cm}\left(H_{1}(x)=-\mu^{2}\right)\nn \\
if\hspace{0.5cm} x \rightarrow +\infty
\hspace{0.2cm} \left(X \rightarrow +\infty \right)
&then\hspace{0.5cm} H_{2}(x)=\frac{\lambda^{2}}{24\pi}\mu^{2}
\hspace{0.2cm}\left(H_{1}(x)=0\right)\nn .
\end{array}
\end{eqnarray*}
\noindent
Through the use of relations (\ref{trace2}), (\ref{H22}), (\ref{H1})
for the stringy black hole
 background (\ref{element2})-(\ref{roots2})
equation (\ref{generalT}) becomes :
\bea
T^{\mu}_{\nu}= & \left[
\begin{array}{cc}
 \frac{\lambda^{2}}{6\pi}
 \left[\frac{2\left(X(x)+\mu -1\right)+\mu(1-X(x))}{\left(X(x)+1
\right)^{2}}\right]
- \Omega^{-1}(x)\left(\frac{\lambda^{2}}{24\pi}\right)
\left[\mu^{2}+H_{1}(x)\right]
& 0 \\
 0 & \Omega^{-1}(x)\left(\frac{\lambda^{2}}{24\pi}\right)
\left[\mu^{2}+H_{1}(x)\right]
\end{array}
\right]
& \nn \\
 & +\hspace{0.5cm}\Omega^{-1}(x)
\left[ \begin{array}{cc}
-\beta & -\alpha\\
\alpha & \beta
\end{array}
\right]. \label{em7} \eea In order to find the explicit form of
the regularized energy-momentum tensor in the different vacua
considered before we follow the analysis of the previous section.
\newline
\newline
\noindent
{\bf (i) Boulware Vacuum}
\newline
In this vacuum there are no particles detected at infinity
($\mathcal{J}^+$) and the regularized energy momentum tensor
(\ref{em7}) should coincide at infinity with the standard Casimir
energy-momentum tensor (\ref{emb}).
\newline
The constants of integration are :
\be
\beta=\frac{\pi}{24L^{2}}-\frac{\lambda^{2}}{24\pi}
\mu^{2}\\
\hspace{1cm}\alpha =0
\ee
and thus the regularized energy-momentum tensor is :
\bea
T^{(\eta)\mu}_{\nu}= & \left[
\begin{array}{cc}
 \frac{\lambda^{2}}{6\pi}
 \left[\frac{2\left(X(x)+\mu -1\right)+\mu(1-X(x))}{\left(X(x)+1
\right)^{2}}\right]
- \Omega^{-1}(x)\left(\frac{\lambda^{2}}{24\pi}\right)
\left[\mu^{2}+H_{1}(x)\right]
& 0 \\
 0 & \Omega^{-1}(x)\left(\frac{\lambda^{2}}{24\pi}\right)
\left[\mu^{2}+H_{1}(x)\right]
\end{array}
\right]
& \nn \\
 & +\hspace{0.5cm}\Omega^{-1}(x)\left(\frac{\pi}{24L^{2}}
-\frac{\lambda^{2}}{24\pi}\mu^{2}\right)
\left[ \begin{array}{cc}
-1 & 0\\
0 & 1
\end{array}
\right]. \label{em8} \eea The detected energy density, pressure
and energy are asymptotically ($x\rightarrow +\infty$, or equivalently
setting $\Omega(x)=1$) given by : \bea
\rho=T^{(\eta)t}_{t}=-\frac{\pi}{24L^{2}}\label{density2}\\
p=-T^{(\eta)x}_{x}=-\frac{\pi}{24L^{2}}\\ E(L) =\int^{L}_{0}\rho
dx= -\frac{\pi}{24L}.\label{energy2}  \eea
 The corresponding force between the ``walls''
is attractive as expected :
\be
F(L)=-\frac{\partial E(L)}{\partial L}=-\frac{\pi}{24L^{2}}< 0  .
\ee In analogy with (\ref{energy11}) the energy for arbitrary
position of the ``walls'' ($x_1$, $x_1 +L$) can be evaluated. The
corresponding expression is quite complicated and is not presented
here but gives (\ref{energy2}) as $x_1 \rightarrow +\infty$.
\newline
\newline
\noindent
{\bf (ii) Hartle-Hawking Vacuum}
\newline
In this vacuum the stringy black hole
(\ref{metric2})-(\ref{roots2})
is in thermal equilibrium with an infinite reservoir
of black body radiation at temperature $T$ which
is equal to the Hawking temperature of the stringy
two-dimensional charged black hole \cite{diamandis} :
\be
T=T_{H}=\frac{\lambda}{2\pi}\mu
\label{temp}.
\ee
\newline
The regularized energy-momentum tensor (\ref{em7}) should coincide
with the modified Casimir energy-momentum tensor (\ref{emb+bath}).
\newline
The constants of integration are :
\be
\beta=\frac{\pi}{24L^{2}} \\ \hspace{1cm}\alpha =0 \ee and thus
the regularized energy-momentum tensor is : \bea
T^{(\upsilon)\mu}_{\nu}= & \left[
\begin{array}{cc}
 \frac{\lambda^{2}}{6\pi}
 \left[\frac{2\left(X(x)+\mu -1\right)+\mu(1-X(x))}{\left(X(x)+1
\right)^{2}}\right]
- \Omega^{-1}(x)\left(\frac{\lambda^{2}}{24\pi}\right)
\left[\mu^{2}+H_{1}(x)\right]
& 0 \\
 0 & \Omega^{-1}(x)\left(\frac{\lambda^{2}}{24\pi}\right)
\left[\mu^{2}+H_{1}(x)\right]
\end{array}
\right]
& \nn \\
 & +\hspace{0.5cm}\Omega^{-1}(x)\left(\frac{\pi}{24L^{2}}
\right)
\left[ \begin{array}{cc}
-1 & 0\\
0 & 1
\end{array}
\right].& \label{em9} \eea The detected energy density, pressure and energy are 
asymptotically ($x\rightarrow +\infty$, or
equivalently setting $\Omega(x)=1$)
given by : \bea
\rho=T^{(\upsilon)t}_{t}=-\left(\frac{\pi}{24L^{2}}
+\frac{\lambda^{2}}{24\pi}\mu^{2}\right)\\
p=-T^{(\upsilon)x}_{x}=-\left(\frac{\pi}{24L^{2}}
+\frac{\lambda^{2}}{24\pi}\mu^{2}\right) \eea
\be
E(L, T_{H})=\int^{L}_{0}\rho dx =-\left(\frac{\pi}{24L}
+\frac{\lambda^{2}}{24 \pi} \mu^{2} L \right)=
-\left(\frac{\pi}{24L} +\frac{\pi L}{6} T^{2}_{H} \right) .
\label{energyHH2} \ee
The corresponding force between the ``walls'' is not always attractive :
\be
F(L, T_{H})=-\left(\frac{\partial E(L)}{\partial
L}\right)_{T_{H}}=-\frac{\pi}{24L^{2}}+
\frac{\lambda^{2}}{24\pi}\mu^{2}=-\frac{\pi}{24L^{2}}+
\frac{\pi}{6}T^{2}_{H} .
 \ee It is obvious again that the corresponding force is :
\begin{description}
\item \hspace{2cm}{\bf (a) attractive}
\be L < \frac{1}{2\hspace{0.1cm} T_{H}}=\frac{\pi}{\lambda \mu}
\ee
\item \hspace{2cm}{\bf (b) zero} \be L = \frac{1}{2\hspace{0.1cm}
 T_{H}} \ee
\item \hspace{2cm}{\bf (c) repulsive}
\be L > \frac{1}{2\hspace{0.1cm} T_{H}} .\ee
\end{description}
Thus as in the case of two-dimensional ``Schwarzschild'' black
hole if the last condition is satisfied the outer ``wall'' moves
towards infinity. It can be studied as a ``moving mirror''
creating particles whose energy rate detected at infinity is given
by the second term in equation (\ref{energyHH2}):
\be
\frac{dE}{dt}=\frac{\lambda^{2}}{24 \pi} \mu^{2} L=\frac{\pi L}{6}
T^{2}_{H} . \ee
This is the rate at which energy is radiated for the case of the
massless two-dimensional field.
\newline
\par
\noindent
{\bf (iii) Unruh Vacuum}
\newline
In this vacuum an outward flux of radiation is detected
at infinity. Thus the stringy two-dimensional charged black hole
(\ref{metric2})-(\ref{roots2}) radiates and its temperature when
the system has settled down to an ``equilibrium'' state is given as in
(\ref{temp}) \cite{diamandis}.
The regularized energy-momentum
tensor (\ref{em7}) should now coincide at infinity with the modified Casimir
energy-momenum tensor (\ref{emb+rad}).
\newline
The constants of integration are :
\be
\beta=\frac{\pi}{24L^{2}}-\frac{\lambda^2}{48\pi}\mu^2\\
\hspace{1cm}\alpha =\frac{\lambda^2}{48\pi}\mu^2
\ee
thus the regularized energy-momentum tensor is now given by :
\bea
T^{(\xi)\mu}_{\nu}= & \left[
\begin{array}{cc}
 \frac{\lambda^{2}}{6\pi}
 \left[\frac{2\left(X(x)+\mu -1\right)+\mu(1-X(x))}{\left(X(x)+1
\right)^{2}}\right]
- \Omega^{-1}(x)\left(\frac{\lambda^{2}}{24\pi}\right)
\left[\mu^{2}+H_{1}(x)\right]
& 0 \\
 0 & \Omega^{-1}(x)\left(\frac{\lambda^{2}}{24\pi}\right)
\left[\mu^{2}+H_{1}(x)\right]
\end{array}
\right]
& \nn \\
 & +\hspace{0.5cm}\Omega^{-1}(x)
\left[
\begin{array}{cc}
-\frac{\pi}{24L^{2}}
+\frac{\lambda^{2}}{48\pi}\mu^{2}
& -\frac{\lambda^{2}}{48\pi}\mu^{2}
\\
\frac{\lambda^{2}}{48\pi}\mu^{2}& \frac{\pi}{24L^{2}}
-\frac{\lambda^{2}}{48\pi}\mu^{2}
\end{array}
\right] .&
\label{em9}
\eea The detected energy density, pressure and energy
are asymptotically ($x\rightarrow +\infty$, or
equivalently setting $\Omega(x)=1$)
given by : \bea \rho=T^{(\xi)t}_{t}=-\left(\frac{\pi}{24L^{2}}
+\frac{\lambda^{2}}{48\pi}\mu^{2}\right)\\
p=-T^{(\xi)x}_{x}=-\left(\frac{\pi}{24L^{2}}
+\frac{\lambda^{2}}{48\pi}\mu^{2}\right) \eea
\be
E(L, T_{H})=\int^{L}_{0}\rho dx = -\left(\frac{\pi}{24L}
+\frac{\lambda^{2}}{48\pi}\mu^{2}L\right)= -\left(\frac{\pi}{24L}
+\frac{\pi L}{12}T^{2}_{H}\right) . \label{energyU2} \ee
The corresponding force
between the ``walls'' is not always attractive :
\be
F(L, T_{H})=-\left(\frac{\partial E(L, T_{H})}{\partial
L}\right)_{T_{H}}=-\frac{\pi}{24L^{2}}+
\frac{\lambda^{2}}{48\pi}\mu^{2}=-\frac{\pi}{24L^{2}}+
\frac{\pi}{12}T^{2}_{H} \ee  and is thus :
\begin{description}
\item \hspace{2cm}{\bf (a) attractive}
\be L < \frac{1}{\sqrt{2} \hspace{0.1cm}
T_{H}}=\sqrt{2}\frac{\pi}{\lambda \mu} \ee
\item \hspace{2cm}\bf (b) zero \be L = \frac{1}{\sqrt{2}\hspace{0.1cm}
 T_{H}} \ee
\item \hspace{2cm}\bf (c) repulsive
\be L > \frac{1}{\sqrt{2}\hspace{0.1cm} T_{H}} . \ee
\end{description}
Thus if the outer ``wall'' is placed at a distance $L$ such that
the last condition is satisfied then it will move
towards infinity. It can be studied as a ``moving mirror''
creating particles whose energy rate detected at infinity is
given by the second term in equation (\ref{energyU2}) :
\be
\frac{dE}{dt}=\frac{\lambda^{2}}{48\pi}\mu^{2}L=
\frac{\pi L}{12}T^{2}_{H}
\ee
which is for the two-dimensional massless field
the rate which the energy is radiated.
\par
\noindent
It is interesting to evaluate in this vacuum some
thermodynamical quantities and to examine these results
with respect to the laws of thermodynamics.
\par \noindent
The specific heat is given as :
\be
C_{V}=-\frac{\partial E}{\partial T} =-\left(\frac{\pi
L}{6}\right) T_{H}= -\left(\frac{L}{12}\right)\lambda\mu \ee and
the isothermal compressibility is :
\be
\kappa_{T}=-\frac{1}{L}\left( \frac{\partial L}{\partial
p}\right)_{T} =-\left(\frac{12}{\pi}\right)L^{2} . \ee The
comments made in the corresponding ``Schwarzschild'' case also hold here.
\par
\noindent
The entropy of the stringy two-dimensional charged
black hole seen as a Casimir system using the first
 thermodynamical law is given by :
\be
S_{Casimir}=S_{\left(extremal\right)}-\left(\frac{\pi L}{6}\right)
 T_{H}=
S_{\left(extremal\right)}-\left(\frac{L}{12}\right)\lambda \mu
\label{entropy2} \ee where $S_{\left(extremal\right)}$ is the
entropy of the two-dimensional extremal ($Q=M$ or equivalently
$\mu=0$) black hole due to vacuum polarization.
\newline
The two-dimensional extremal black hole is shown \cite{diamandis} to be obtained as
a regular limit of the stringy two-dimensional charged (nonextremal)
black hole. The results obtained in the case of the
stringy two-dimensional ``Schwarzschild'' black hole are those
of the stringy two-dimensional charged black hole when the
parameter $\mu$ approaches $1$, i.e. the electric charge is zero
($Q=0$) \cite{diamandis}. Therefore we expect to get equation (\ref{entropy1})
by setting $\mu=1$ to equation (\ref{entropy2}).
In order to achieve this we must set
the entropy of the extremal black hole
$S_{\left(extremal\right)}$ due to vacuum polarization
 equal to zero . The thermodynamical entropy of the
two-dimensional charged black hole is given now by :
\be
S_{Casimir}=-\left(\frac{\pi L}{6}\right)
 T_{H}=
-\left(\frac{L}{12}\right)\lambda \mu . \label{entropy3} \ee
This is the part of the thermodynamical entropy due to the vacuum polarization
as mentioned before.
The statistical-mechanical part $S^{SM}$ of the thermodynamical entropy
is absent and since the entropy calculated here does not have a statistical
interpretation is not prohibited to be negative.

\section{Discussion}
In this paper we have explicitly calculated in the stringy two-dimensional
``Schwarzschild'' and  ``Reissner-Nordstr\"{o}m'' black hole backgrounds the regularized
energy-momentum tensor of a massless scalar field satisfying the Dirichlet
boundary conditions. The
regularized energy-momentum tensor is separately treated in the
Boulware, Hartle-Hawking and Unruh vacua. In the Boulware vacuum
the asymptotically detected energy, energy  density, pressure acting on
the ``walls'' and the corresponding force
between the ``walls'' where proved to be the same for both stringy
black hole backgrounds (``Schwarzschild'' and ``Reissner-Nordstr\"{o}m'').
In the other two vacua the expressions obtained for the ``Reissner-Nordstr\"{o}m"
 black hole are seen to approach the
corresponding results for the ``Schwarzschild'' case when $\mu \rightarrow 1$, i.e.
the electric charge is zero ($Q=0$). We have shown that the corresponding force
between the ``Dirichlet walls'' is not always attractive :
it can be attractive, zero or repulsive depending on the
distance between the ``walls'' being smaller, equal or
larger of the inverse Hawking temperature  of the black hole. This
can be understood if we recall the semi-infinite Witten's cigar
which is an interpretation of an Euclidean black hole and which is
asymptotic to a cylindrical two-dimensional spacetime. The inverse
temperature of the Euclidean black hole can be viewed as the
circumference of the Witten's cigar. Therefore if the distance
between the ``Dirichlet walls'' is smaller than the circumference
of the Witten's cigar then the corresponding force will be dominated by the
attractive term due to the boundary effects and not by the
repulsive term of the radiation pressure. If the ``Dirichlet
walls'' are placed at a distance approximately equal to the
circumference of the Witten's cigar the attractive term will be
compensated by the repulsive term and the corresponding force will be zero.
Finally if the ``Dirichlet walls'' are placed at a distance larger than
the circumference then the dominant term will be the repulsive
term due to the radiation pressure and the corresponding force acting on the
``walls'' will be repulsive.
\par \noindent
We have concluded that the thermodynamical quantities specific
heat and thermal compressibility violate the second
thermodynamical law since they obtain negative values. The
thermodynamical second law instability induced by the specific
heat is not suprising since it is a byproduct of the semiclassical
approximation (one-loop gravity). This is the reason we reach the
same result in the four-dimensional black hole physics. The
thermodynamical stability is regained by ``freezing'' the black
hole which in the case of the two-dimensional
 charged black hole means to reach extremality ($Q=M$). The
thermodynamical second law instability induced by the negativity
of the isothermal compressibility reflects the possibility of
extracting energy from the vacuum. The thermodynamical entropy was
evaluated in the Unruh vacuum and was shown to be negative. This
is not a violation of the second thermodynamical law since the
thermodynamical (Casimir) entropy calculated here - the
statistical-mechanical part of entropy is absent - is due to the
vacuum polarization. Thus we have obtained the one-loop correction
(due to vacuum polarization) to the classical entropy obtained
from the corresponding classical gravitational action. Same result
can be easily obtained for the case of Hartle-Hawking vacumm in
contradistinction to the Casimir entropy of the Boulware vacuum
which is zero since energy is temperature independent.

 \par \noindent In the extremal case ($Q=M$) of the
stringy two-dimensional charged black hole the results obtained in
the Hartle-Hawking and Unruh vacua coincide with those of stable
Boulware vacuum. This is another argument which strengthens our
belief that extremal black holes are a stable quantum mechanical
ending point for the black holes in the process of their
evaporation.
\par \noindent Finally we would also like to note that, 
for asymptotic position of the ``walls'',  it is natural 
for the results to depend only on the temperatute of 
the black hole; however for arbitrary
position of the ``walls'' the dependence of the energy on the
position is different for the two stringy two-dimensional black
holes studied in this work. Possible determination (using this difference) 
of the nature of a black hole through the Casimir effect 
is very interesting and we have in mind to come back with a future work.

\section*{Acknowledgements}
The authors acknowledge partial financial support by the University
of Athens' Special Account for the Research.


\end{document}